\documentclass[a4paper,prl,superscriptaddress,twocolumn]{revtex4}
\usepackage[dvips]{graphicx}
\usepackage{amssymb}
\newcommand{\ped}[1]{\ensuremath{_{\rm #1}}}
\newcommand{\apex}[1]{\ensuremath{^{\rm #1}}}

\begin{document}

\title{The effect of magnetic impurities in a two-band superconductor:\\
A point-contact study of Mn-substituted MgB$_2$ single crystals}

\author{R.S.~Gonnelli\email{E-mail:renato.gonnelli@polito.it}}
\author{D.~Daghero}
\author{G.A.~Ummarino}
\author{A.~Calzolari}
\author{M.~Tortello}
\affiliation{Dipartimento di Fisica and CNISM, Politecnico di
Torino, 10129 Torino, Italy}
\author{V.A.~Stepanov}
\affiliation{P.N. Lebedev Physical Institute, Russian Academy of
Sciences, 119991 Moscow, Russia}
\author{N.D.~Zhigadlo}
\author{K. Rogacki}
\altaffiliation{\vspace{-5mm}On leave from: Institute of Low
Temperature and Structure Research, Polish Academy of Sciences,
50-950 Wroclaw, Poland.}
\author{J.~Karpinski}
\affiliation{Laboratory for Solid State Physics, ETHZ, CH-8093
Zurich, Switzerland}

\author{F.~Bernardini}
\author{S. Massidda}
\affiliation{Dipartimento di Fisica, Universit\`{a} di Cagliari,
09042 Monserrato (CA), Italy}


\begin{abstract}
We present the first results of directional point-contact
measurements in Mg\ped{1-x}Mn\ped{x}B\ped{2} single crystals, with
$x$ up to 0.015 and bulk $T\ped{c}$ down to 13.3 K. The order
parameters $\Delta\ped{\sigma}$ and $\Delta\ped{\pi}$ were
obtained by fitting the conductance curves with the two-band
Blonder-Tinkham-Klapwijk (BTK) model. Both $\Delta\ped{\pi}$ and
$\Delta\ped{\sigma}$ decrease with the critical temperature of the
junctions $T\ped{c}\apex{A}$, but remain clearly distinct up to
the highest Mn content.  Once analyzed within the Eliashberg
theory, the results indicate that spin-flip scattering is dominant
in the $\sigma$ band, as also confirmed by first-principle band
structure calculations.
\end{abstract}
\maketitle

%\section{Introduction}
The two-band character of superconductivity in MgB$_2$
\cite{Liu,Brinkman} has been almost completely understood and
explained by now, but the effects of disorder and chemical doping
are still in need of some experimental clarification. Several
substitutions have been tried, either in the Mg or in the B sites,
but only few of them have been successful \cite{Cava} and none has
been able to enhance the critical temperature of the compound.
However, their study has proven useful to clarify the role of the
different scattering channels and to try to control them
selectively. Substitutions with magnetic impurities (Mn, Fe)
represent a class of its own because of the spin-flip
pair-breaking scattering that is expected to dramatically suppress
superconductivity -- even though the way it does it in a two-band
superconductor has never been studied experimentally. Up to now,
only two successful Mn substitution in the Mg site have been
reported, one in polycrystalline samples \cite{Xu} and one in
single crystals \cite{Karpinski_Mn}. Here, we present the first
results of point-contact spectroscopy (PCS) measurements in
Mn-substituted MgB$_2$ single crystals grown at ETHZ, which
allowed us to study the effects of magnetic impurities on the
order parameters (OPs) of a two-band superconductor. The
amplitudes of the OPs, $\Delta\ped{\sigma}$ and $\Delta\ped{\pi}$,
were determined as a function of the critical temperature. The
resulting trend can be explained within the Eliashberg theory as
being mainly due to a doping-induced increase in the pair-breaking
scattering within the $\sigma$ bands, with minor contributions
from the $\pi$-$\pi$ or the $\sigma$-$\pi$ channels. This result,
apparently in contrast with the Mn position in the lattice, is
however confirmed by first-principle calculations of the local
effect of a Mn impurity on the bandstructure of MgB$_2$.

%\section{Experimental details}
The high-quality Mg\ped{1-x}Mn\ped{x}B\ped{2} single crystals used
for our measurements were grown by using the same high-pressure,
cubic-anvil technique set up for pure MgB$_2$, and by replacing part
of the Mg precursor with metallic Mn \cite{Karpinski_Mn}. The Mn
content $x$ of \emph{each crystal} was measured by EDX through a
careful evaluation of the Mn/Mg ratio; the crystals are single-phase
and homogeneous within $\pm\delta x$ ($\delta x$=0.0010). The bulk
critical temperature T\ped{c} was determined by DC magnetization
measurements. As reported elsewhere \cite{Karpinski_Mn}, the
dependence of the lattice constants on the Mn content indicates that
Mn replaces Mg in the lattice. The magnetic-field dependence of the
Curie part C\apex{*} of the magnetic moment M clearly indicates that
Mn ions are divalent (i.e. Mn$^{2+}$) and in the low-spin state
($S=1/2$), as also confirmed by our first-principle band-structure
calculations. The crystals we used were carefully selected among
those with the sharpest superconducting transitions and best
structural properties, so that secondary phases or other impurities
can be excluded. They had different Mn contents $x$ between 0.0037
and 0.0150, corresponding to bulk critical temperatures $T\ped{c}$
between 33.9 and 13.3~K, with $\Delta T\ped{c}(10-90\%)$ increasing
with doping from 0.65 K to 5.4 K. The point contacts were made by
putting a small drop of silver paint ($\varnothing\simeq 50 \mu$m)
on the flat side surface of the crystal~\cite{nostroPRL} so as to
inject the current mainly parallel to the $ab$ planes, which in pure
MgB$_2$ is the best configuration for the observation of both the
$\sigma$- and $\pi$-band gaps \cite{Brinkman,nostroPRL}. In most
cases, we studied the temperature dependence and the magnetic-field
dependence of the conductance curves (d$I$/d$V$ vs. $V$), so as to
determine the critical temperature of the junction (i.e., the
``Andreev critical temperature'', $T\ped{c}\apex{A}$), and to
understand whether one or two OPs were present
\cite{nostroPRL,nostroPRB}. The conductance curves were then
normalized to the normal state and fitted with the two-band BTK
model \cite{BTK}, as described elsewhere \cite{nostroPRL}. The
fitting function contains the OPs $\Delta\ped{\sigma,\pi}$, the
coefficients $Z\ped{\sigma,\pi}$ (related to the potential barrier
and to the the Fermi velocity mismatch at the interface) and the
lifetime broadening coefficients $\Gamma\ped{\sigma, \pi}$ as
adjustable parameters, plus the weight $w\ped{\pi}$ of the
$\pi$-band contribution to the conductance. For reliable estimates
of the OP amplitudes, whenever possible, we selected contacts with
rather high resistance ($R\ped{N}\gtrsim 30 \,\Omega$), and with no
dips \cite{dips} in the conductance curves, so as to fulfill the
requirements for ballistic conduction \cite{Duif}.

%%%%%%%%%%%%%%%%%%%%%%%%%%%%%%%%%%%%%%%%%%%%%%%%%%%%%%%%%%%%%%%%%%%%
\begin{figure}[t]
\begin{center}
\includegraphics[keepaspectratio, width=0.7\columnwidth]{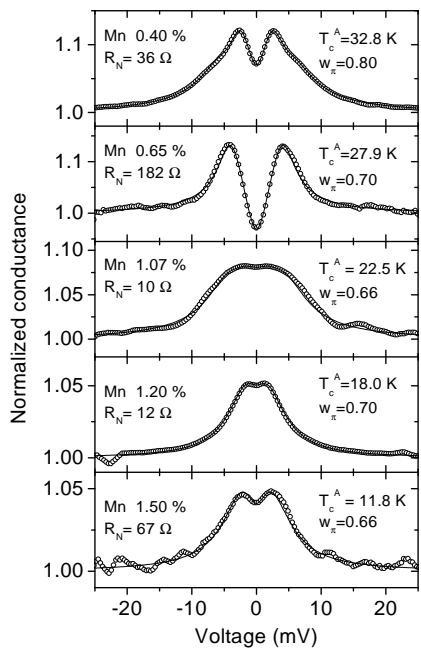}
\end{center}
\vspace{-5mm} %
\caption{\small{Normalized conductance curves (symbols) of
different \emph{ab}-plane junctions in
Mg\ped{1-x}Mn\ped{x}B\ped{2} single crystals at T=4.2 K. The
curves are labeled with the corresponding $T\ped{c}\apex{A}$
values. The Mn content and the normal-state junction resistance
are shown as well. Lines: best-fit curves given by the two-band
BTK model. The $\pi$-band weight $w\ped{\pi}$ is also indicated.
In the absence of specific theoretical predictions, we took $0.66
\le w\ped{\pi} \le 0.8$ as for $ab$-plane contacts in pure MgB$_2$
\cite{Brinkman,nostroPRL}.}} \label{Fig:totale} \vspace{-5mm}
\end{figure}
%%%%%%%%%%%%%%%%%%%%%%%%%%%%%%%%%%%%%%%%%%%%%%%%%%%%%%%%%%%%%%%%%%%%%

Fig.~\ref{Fig:totale} reports some normalized conductance curves
(symbols) in crystals with different Mn content. From now on, we
will label the curves with the corresponding value of
$T\ped{c}\apex{A}$ instead of the Mn content or the bulk
$T\ped{c}$, since PCS is a local, surface-sensitive probe.
$T\ped{c}\apex{A}$ can be smaller than $T\ped{c}$ if the bulk
transition is broad (as for the most-doped crystals) or if
proximity effect occurs at the S/N interface (when $\xi
\thickapprox a$, being $a$ the true contact size).

Lines in Fig.~\ref{Fig:totale} represent the two-band BTK fit of
experimental curves. In conventional superconductors, spin-flip
scattering makes the superconducting gap become ill-defined
\cite{AbrikosovGorkov} and, for strong scattering, bands of states
within the original energy gap are formed \cite{Shiba,Carbotte}.
Strong-coupling calculations for MgB$_2$ \cite{Moca} have shown
that magnetic impurities can give rise to sub-gap states in both
the partial $\sigma$- and $\pi$-band DOS, that are not taken into
account by the BTK model. However, the sub-gap features in the DOS
are by far smaller than the peaks connected to the OPs
$\Delta\ped{\pi}$ and $\Delta\ped{\sigma}$ \cite{Moca}. At any
finite temperature they are further smeared out, so that they
would be very difficult to observe experimentally. One can
reasonably expect that the same happens in the Andreev-reflection
curves. As a matter of fact, all the experimental conductance
curves are rather broadened and have smaller amplitude (see
Fig.\ref{Fig:totale}) than in pure MgB$_2$ \cite{nostroPRL}, so
that even if sub-gap structures exist, they are practically
unobservable. Hence, the BTK model can be used as a reasonable,
first-order approximation to a more specific model for Andreev
reflection in the presence of magnetic impurities on the
superconducting side of the contact, which is lacking at the
present moment.

%%%%%%%%%%%%%%%%%%%%%%%%%%%%%%%%%%%%%%%%%%%%%%%%%%%%%%%%%%%%%%%%%%%%
\begin{figure}[t]
\vspace{-1mm}
\begin{center}
\includegraphics[keepaspectratio, width=0.7\columnwidth]{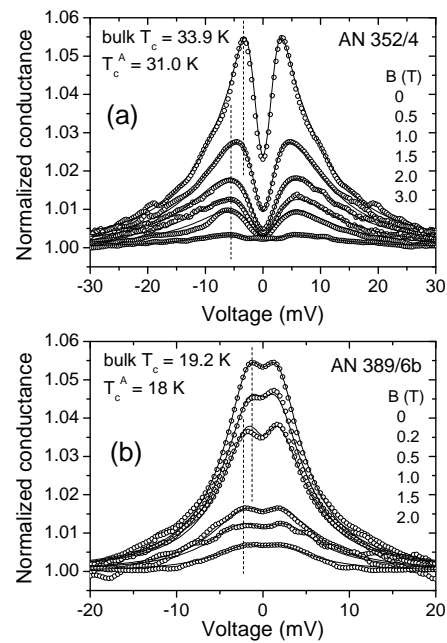}
\end{center}
\vspace{-5mm} %
\caption{\small{Experimental conductance curves at T=4.2~K of two
contacts having $T\ped{c}\apex{A}$=31~K (a) and
$T\ped{c}\apex{A}$=18~K (b), in the presence of different magnetic
fields $\mathbf{B}\parallel c$ (circles) and the relevant two-band
BTK fit (lines). Vertical dotted lines indicate the position of
the conductance peaks. The values of the fitting parameters in
zero field are: $\Delta\ped{\pi}$=2.5~meV,
$\Delta\ped{\sigma}$=5.2~meV; $\Gamma\ped{\pi}$=2.3~meV,
$\Gamma\ped{\sigma}$=6.0~meV; $Z\ped{\pi}$=0.55,
$Z\ped{\sigma}$=0.82, $w\ped{\pi}$=0.7 for panel (a);
$\Delta\ped{\pi}$=1.56~meV, $\Delta\ped{\sigma}$=2.8~meV;
$\Gamma\ped{\pi}$=2.01~meV, $\Gamma\ped{\sigma}$=3.8~meV;
$Z\ped{\pi}$=0.43, $Z\ped{\sigma}$=0.50, $w\ped{\pi}$=0.7 for
panel (b). The two-band BTK fit works well up to the critical
field $B\ped{c2}$ ($\simeq$ 3.5~T in (a) and $\simeq$ 2.5~T in
(b)).}} \label{fig:Bdep} \vspace{-5mm}
\end{figure}
%%%%%%%%%%%%%%%%%%%%%%%%%%%%%%%%%%%%%%%%%%%%%%%%%%%%%%%%%%%%%%%%%%%%%

Even if the fit of the conductance curves shown in
Fig.~\ref{Fig:totale} indicates the existence of two OPs in the
Mg$\ped{1-x}$Mn$\ped{x}$B$_2$ system, it is clear that some curves
-- especially at the highest doping levels -- show little or no
structures associated to the larger OP. To check in a more
convincing way if two-band superconductivity persists up to the
highest Mn content, we studied the magnetic-field dependence of
the d$I$/d$V$ curves in the whole range of $T\ped{c}\apex{A}$.

%%%%%%%%%%%%%%%%%%%%%%%%%%%%%%%%%%%%%%%%%%%%%%%%%%%%%%%%%%%%%%%%%%%%
\begin{figure}[t]
\vspace{-2mm}
\begin{center}
\includegraphics[keepaspectratio, width=0.9\columnwidth]{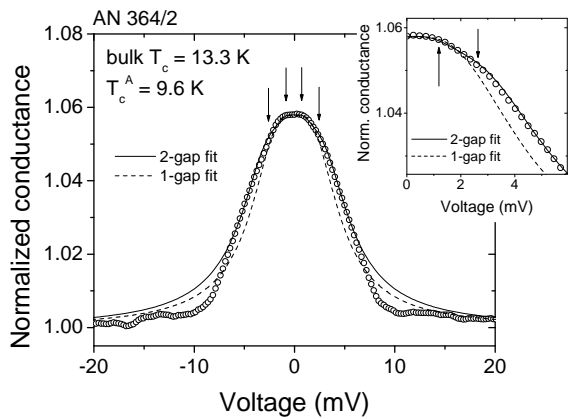}
\end{center}
\vspace{-5mm} %
\caption{\small{Symbols: zero-field, low-temperature (T=4.2 K)
conductance curve of a contact with $T\ped{c}\apex{A}=9.6$~K made
on the most-doped crystals ($x=0.015$). Solid line: best fit given
by the two-band BTK model. Dashed line: best fit given by the
standard (single-band) BTK model. Arrows indicate the smooth
structures related to the two order parameters, which are
magnified in the inset for clarity.}} \label{fig:fit_lowest_Tc}
\vspace{-5mm}
\end{figure}
%%%%%%%%%%%%%%%%%%%%%%%%%%%%%%%%%%%%%%%%%%%%%%%%%%%%%%%%%%%%%%%%%%%%

Fig.~\ref{fig:Bdep}(a) shows the magnetic-field dependence of the
conductance curve of a contact with $T\ped{c}\apex{A}$=31~K
(circles) with the relevant two-band BTK fit (lines). The
zero-field curve shows both the peaks corresponding to
$\Delta\ped{\pi}$ and the smooth shoulders related to
$\Delta\ped{\sigma}$. On applying the magnetic field, the
small-gap features reduce in amplitude, progressively unveiling
the underlying large-gap features. An outward shift of the peaks
is observed at some $B=B\apex{*}$, when the $\sigma$-band
structures become dominant and determine the shape of the
conductance curve. Unlike in pure MgB$_2$, here $B\apex{*}$ is
intense enough (if compared to the critical field, strongly
suppressed by Mn impurities) to partly depress
$\Delta\ped{\sigma}$, while the $\pi$-band partial conductance
seems not to vanish at $B^{*}$. This makes it impossible to
separate the partial $\sigma$- and $\pi$- band conductances as we
did in pure MgB$_2$ \cite{nostroPRL,nostroPRB}, but the shift of
the conductance peaks is anyway a clear proof of the existence of
two OPs. The same happens in the magnetic-field dependence of
contacts with much lower critical temperature. For example, Fig.
\ref{fig:Bdep}(b) reports the field dependence of the normalized
conductance curve of a contact with $T\ped{c}\apex{A}$=18~K. Here
the zero-field conductance does not show clear structures related
to $\Delta\ped{\sigma}$ but an outward shift of the conductance
maxima is observed anyway at $B=B\apex{*}\simeq 1$~T. For higher
doping levels, i.e. for $T\ped{c}\apex{A} < 17$~K, the
low-temperature, zero-field conductance curves can be very well
fitted with the two-band BTK model (see the bottom curve in
Fig.\ref{Fig:totale}), but the critical field is so small and the
conductance curves are so broadened that their magnetic-field
dependence is not conclusive, at least with our experimental
resolution. Moreover, some conductance curves in this region allow
a standard (i.e. single-band) BTK fit as well, which gives an OP
amplitude $\Delta$ that is the ``average'' of the values given by
the two-band fit. However, the single-band fit often fails in
reproducing both the position of the peaks \emph{and} the width of
the Andreev-reflection structures in the conductance, as shown in
Fig.~\ref{fig:fit_lowest_Tc} for a contact with the lowest
$T\ped{c}\apex{A}$. This suggests that two OPs are likely to be
present also in the most doped samples.

%%%%%%%%%%%%%%%%%%%%%%%%%%%%%%%%%%%%%%%%%%%%%%%%%%%%%%%%%%%%%%%%%%%%
\begin{figure}[t]
\begin{center}
\includegraphics[keepaspectratio, width=0.85\columnwidth]{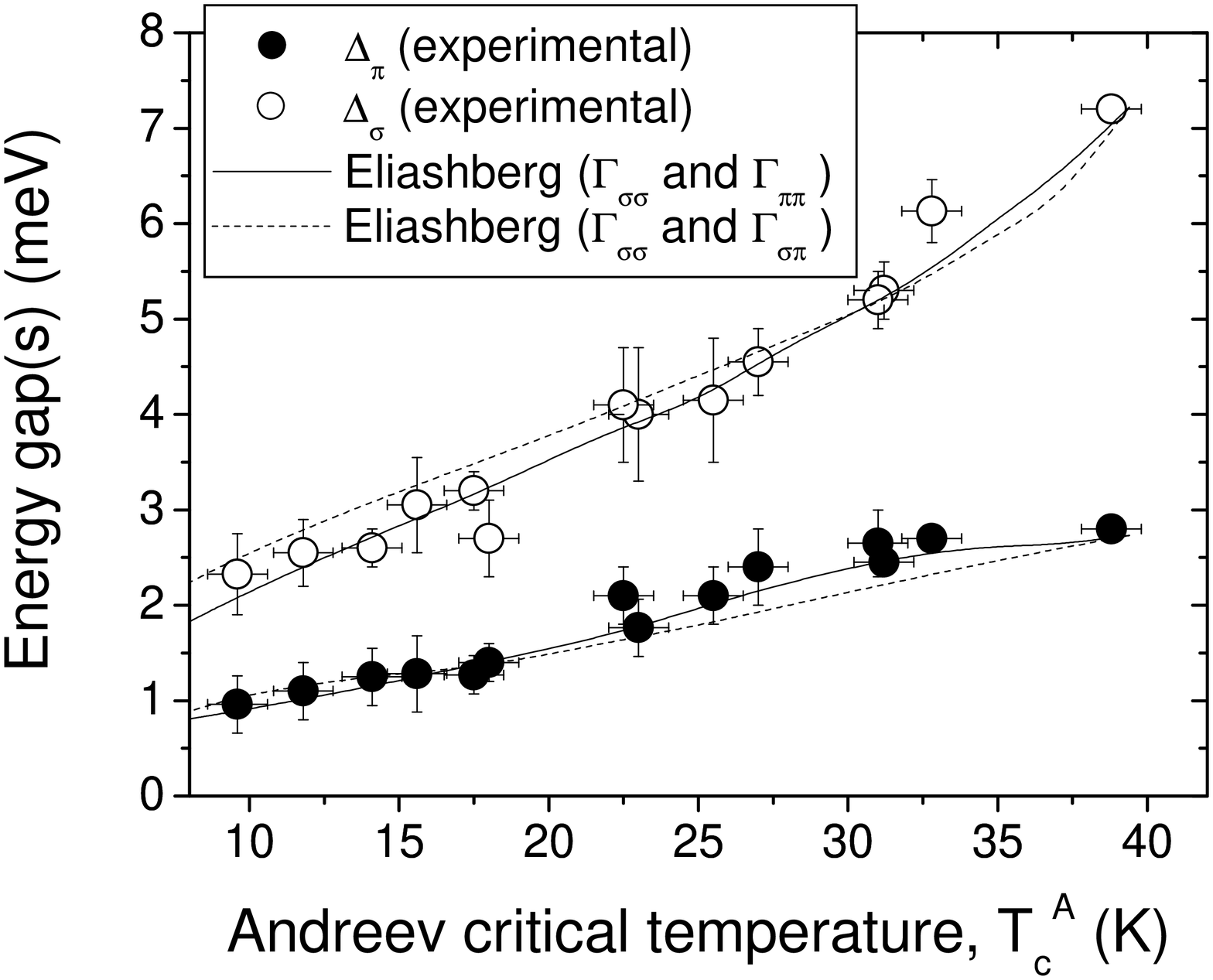}
\end{center}
\vspace{-5mm} %
\caption{\small{Symbols: amplitudes of the order parameters as a
function of the critical temperature of the junctions,
$T\ped{c}\apex{A}$. Lines: theoretical curves obtained by solving
the Eliashberg equations in the presence of magnetic scattering in
the $\sigma$-$\sigma$ and $\pi$-$\pi$ channels (case (a), solid
lines) or in the $\sigma$-$\sigma$ and $\sigma$-$\pi$ channels
(case (b), dashed lines). }} \label{fig:gaps_vs_Tc} \vspace{-5mm}
\end{figure}
%%%%%%%%%%%%%%%%%%%%%%%%%%%%%%%%%%%%%%%%%%%%%%%%%%%%%%%%%%%%%%%%%%%%%

The complete dependence of the OPs on $T\ped{c}\apex{A}$ is
reported in Fig.~\ref{fig:gaps_vs_Tc}. Vertical error bars include
both the uncertainty on the OP in each fit and the spread of OP
values over different contacts with the same $T\ped{c}\apex{A}$
(within the experimental uncertainty on $T\ped{c}\apex{A}$
represented by horizontal error bars). For $T\ped{c}\apex{A} <
17$~K only the results of the two-band BTK fit are shown, based on
the preceding discussion and on the regular trend of the OPs for
$T\ped{c}\apex{A} \geq 18$~K. In fact, the presence of a single OP
in the low-$T\ped{c}\apex{A}$ region would imply a sudden change
in the slope of the curves around $T\ped{c}\apex{A}=17$~K that is
not justified by any observed discontinuity in the physical
properties of the compound \cite{Karpinski_Mn}. The persistence of
two distinct OPs up to the highest $x$ value indicates that Mn
doping does not significantly increase non-spin-flip
\emph{interband} scattering. Actually, this is also suggested by
the very low Mn concentration (1.5 \% at most) and by the decrease
of both the OPs with $T\ped{c}\apex{A}$ (while interband
scattering would increase $\Delta\ped{\pi} $\cite{Kortus}). This
decrease could arise from changes in the DOSs at the Fermi level
\cite{Kortus} and in the phonon spectra $\alpha^2
F\ped{\sigma,\pi}(\omega)$, but these two effects are certainly
negligible here, because Mn is isovalent with Mg
\cite{Karpinski_Mn} and the Mn content is very small. The latter
reason also allows assuming the effect of $\sigma$ and $\pi$
\emph{intraband} non-spin-flip scattering to be much smaller than
that of pair-breaking scattering in the same channels. This leads
to conclude that the main possible cause of the experimental trend
of Fig.~\ref{fig:gaps_vs_Tc} is an increase in the spin-flip
scattering in the $\sigma$-$\sigma$, $\pi$-$\pi$ or $\sigma$-$\pi$
channels.

\begin{figure}[ht]
\vspace{-2mm}
\begin{center}
\includegraphics[keepaspectratio, width=0.9\columnwidth]{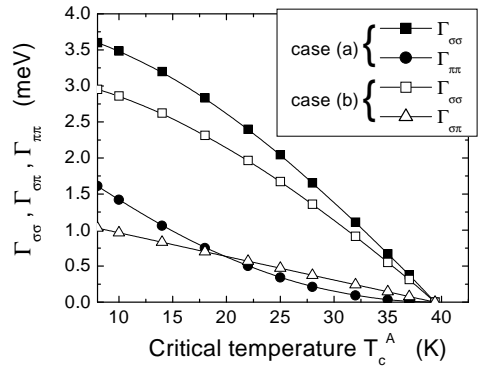}
\end{center}
\vspace{-5mm} %
\caption{\small{Spin-flip scattering parameters used to fit the
data of Fig.\ref{fig:gaps_vs_Tc} as a function of
$T\ped{c}\apex{A}$. Solid symbols: case (a),
$\Gamma\ped{\sigma\sigma}$ (squares) and $\Gamma\ped{\pi \pi}$
(circles). Open symbols: case (b), $\Gamma\ped{\sigma\sigma}$
(squares) and $\Gamma\ped{\sigma \pi}$ (triangles). Lines are
guides to the eye.}} \label{Fig:gamma}
\vspace{-5mm} %
\end{figure}

This simple picture can be made quantitative by solving the
Eliashberg equations (EE) in the presence of randomly distributed
magnetic impurities, treated within the Born approximation
\cite{Dolgov}. According to the reasoning above, we used the same
phonon spectra, DOS values and Coulomb pseudopotential as in pure
MgB$_2$ \cite{Brinkman,nostroPRL}, and neglected all the
non-spin-flip scattering rates.  We thus took as the only
adjustable parameters the \emph{spin-flip} scattering rates
\emph{within} the bands ($\Gamma\ped{\sigma \sigma}$ and
$\Gamma\ped{\pi \pi}$) and \emph{between} bands
($\Gamma\ped{\sigma \pi}$). We immediately found that
$\Gamma\ped{\sigma \sigma}$ is necessary to fit the
$T\ped{c}\apex{A}$ and OP values and that it \emph{must} be
greater than both $\Gamma\ped{\pi \pi}$ and $\Gamma\ped{\sigma
\pi}$ -- otherwise $\Delta\ped{\pi}$ decreases too fast on
decreasing $T\ped{c}\apex{A}$. As for $\Gamma\ped{\sigma \pi}$,
this agrees with the predictions of Ref.~\onlinecite{Moca}. For
simplicity, we analyzed separately the two cases: (a)
$\Gamma\ped{\sigma \sigma}>\Gamma\ped{\pi \pi}$,
$\Gamma\ped{\sigma \pi}$=0 and (b) $\Gamma\ped{\sigma \sigma}>
\Gamma\ped{\sigma \pi}$, $\Gamma\ped{\pi \pi}$=0. First of all, we
fixed $\Gamma\ped{\pi \pi}$=$\Gamma\ped{\sigma
\sigma}$=$\Gamma\ped{\sigma \pi}$=0 at $T\ped{c}\apex{A}$=39.4~K
(i.e. $x$=0). Second, we found the values of the parameters that
give the experimental values of $\Delta\ped{\sigma}$ and
$\Delta\ped{\pi}$ at $T\ped{c}\apex{A}$=18.0~K (cases (a) and
(b)), by solving the imaginary-axis EE and analytically continuing
the solution to the real axis. Once determined the values of
$\Gamma\ped{\pi \pi}$ and $\Gamma\ped{\sigma \pi}$ in these two
points, we searched for the simplest $\Gamma\ped{\pi
\pi}(T\ped{c}\apex{A})$ and $\Gamma\ped{\sigma
\pi}(T\ped{c}\apex{A})$ curves connecting them and allowing the
fit of the experimental values of $T\ped{c}\apex{A}$ and of the
OPs in the whole doping range, with no restrictions on
$\Gamma\ped{\sigma \sigma}$. We found out that these curves are a
parabola (for $\Gamma\ped{\pi \pi}$) and a straight line (for
$\Gamma\ped{\sigma \pi}$). In both cases, $\Gamma\ped{\sigma
\sigma}$ follows an almost parabolic trend as a function of
$T\ped{c}\apex{A}$ (see Fig.\ref{Fig:gamma}). The resulting
theoretical curves that best fit the OPs are reported in
Fig.\ref{fig:gaps_vs_Tc} as solid lines (case (a)) and dashed
lines (case (b)). The agreement between experimental data and
theoretical calculations (especially in case (a)) is striking in
the whole range of $T\ped{c}\apex{A}$. That our simple -- but
reasonable -- model works so well suggests that, up to $x=0.015$,
Mg$\ped{1-x}$Mn$\ped{x}$B$_{2}$ can be treated as a perturbation
of the unsubstituted compound, with only the addition of $\sigma$
intraband magnetic scattering and minor contributions from either
the $\pi$ intraband or the interband scattering.

The intense Mn pair-breaking scattering in the $\sigma$-$\sigma$
channel has been recently predicted as being due to the
hybridization of the $\sigma$ bands of MgB$_2$ with the $d$ orbitals
of Mn \cite{Singh}. To demonstrate that scattering in this channel
is much greater than in the $\pi$-$\pi$ and $\sigma$-$\pi$ ones, we
performed preliminary calculations of the electronic structure near
a Mn substitutional impurity in a $2 \times 2\times 4$ MgB$_2$
superlattice. The results show that around the Fermi level ($E_F$)
there is a spin-down $d_{z^2}$ band, responsible for the Mn magnetic
moment, very sensitive to the details of structural parameters
\cite{Singh}. The $\sigma$ and $\pi$ bands behave quite differently
near the Mn impurity: while the $\pi$ electron spectral density is
depleted around $E_F$ (due to the $\pi$-$d$ interaction), the
$\sigma$ bands have a large amplitude. Furthermore, the $\sigma$
bands show a sizeable ($\geq$ 15~meV) exchange splitting near $E_F$,
larger than the superconducting gap -- which is consistent with the
experimentally observed complete suppression of superconductivity at
about 2\% of Mn \cite{Karpinski_Mn}. On the other hand, the
impurity-induced mixing of $\sigma$ and $\pi$ states around Mn
(providing a qualitative indication of interband scattering), is
present but not very important. These results can explain the quick
drop of $T_c$, with the persistence of the two distinct gaps, and a
larger scattering within the $\sigma$ band \cite{Bernardini}.

%%%%%%%%%%%%%%%%%%%%%%%%%%%%%%%%%%%%%%%%%%%%%%%%%%%%%%%%%%%%%%%%%%%%
%%%%%%%%%%%%%%%%%%%%%%%%%%%%%%%%%%%%%%%%%%%%%%%%%%%%%%%%%%%%%%%%%%%%%

In conclusion, we have presented the results of the first
experimental study of the effects of magnetic impurities on the
order parameters $\Delta\ped{\sigma}$ and $\Delta\ped{\pi}$ of a
two-band superconductor. We have shown that, in
Mg\ped{1-x}Mn\ped{x}B\ped{2}, $\Delta\ped{\sigma}$ and
$\Delta\ped{\pi}$ decrease regularly with the critical temperature
but remain clearly distinct down to the lowest $T\ped{c}\apex{A}$.
Within the Eliashberg theory, this is due to an increase in
spin-flip scattering in the $\sigma$ bands on increasing the Mn
content, with possible minor contributions from the $\sigma$-$\pi$
or the $\pi$-$\pi$ channels. This somehow unexpected conclusion is
also supported by first-principle calculations of the
bandstructure of MgB$_2$ in the vicinity of a Mn impurity.

This work was done within the PRIN Project N. 2004022024 and the
INTAS Project N. 01-0617. V.A.S. acknowledges support by RFBR (Proj.
N. 06-02-16490).

\end{document}